\documentstyle[11pt,twoside,newpasp,epsf]{article}

\begin{document}

\title{Optical and Near-IR Monitoring of Symbiotic Binary Systems}

\author{Joanna Miko{\l}ajewska}
\affil{N. Copernicus Astronomical Center, Bartycka 18,  00716 Warsaw, Poland, e-mail: 
mikolaj@camk.edu.pl}

\begin{abstract} Symbiotic stars are long-period interacting binary systems in which an 
evolved red giant star transfers material to its much hotter compact  companion. Such a 
composition places them among the most variable stars. In addition to periodic variations 
due to the binary motion they often show irregular changes due to nova-like eruptions of 
the hot component. In some systems the cool giant is a pulsating Mira-type star usually 
surrounded by a variable dust shell. Here, I present results of optical and IR monitoring 
of symbiotic systems as well as future prospects for such studies. \end{abstract}

\section{Introduction} 

Most stars in the Universe are binaries. Among them, symbiotic 
stars are interacting binaries in which  an evolved giant transfers material to a hot 
and compact companion. In a typical configuration, a symbiotic binary contains an M\,III 
giant and a white dwarf accreting material lost in the cool giant  wind. The wind is 
ionised by the hotter of the binary components giving rise to symbiotic nebula (cf. 
Miko{\l}ajewska 1997).

Based on the near-IR colours  two distinct classes of symbiotic stars were defined (Allen 
1983):  the S-type (stellar) with normal red giants, and the D-type (dusty) with Mira 
primaries surrounded by a warm dust shell. The distinction between S and D types seems to 
be one of orbital separation: the binary must have enough space for the red giant, and 
yet allow it to transfer sufficient mass to its companion. In fact, all symbiotic systems 
with known orbital periods -- of the order of a few years -- belong to the S-type, while 
the orbital periods for D-type systems are generally not known probably because they are 
longer than periods covered by existing observations (cf. Belczy{\'n}ski et al. 2000). 

Symbiotic stars are thus interacting binaries with the longest orbital periods and the 
largest component separations, and their study is essential to understand the evolution 
and interactions of detached and semi-detached binary stars. They are also among the 
brightest (intrinsically) stars, which makes them excellent observational targets both in 
our Galaxy as well as in nearby galaxies even for relatively small telescopes.

In the following, I will present results of optical and infrared monitoring of symbiotic 
stars as well as future prospects for such studies.

\section{Variable phenomena in symbiotic stars}

The composition of a typical symbiotic binary, specifically  the presence of an evolved 
giant and its accreting companion, places symbiotic stars among the most variable stars. 
They can fluctuate in several different ways, which can be revealed and studied by 
patient monitoring of their light curves and radial velocity changes. Namely,  binary 
motion can be manifested by eclipses of the hot component by the giant, modulation of the 
giant's light due to reflection effect (with orbital period) and due to tidal distortion 
(with $P_{\rm orb}/2$), as well as  radial velocity changes. The cool giant can also show 
intrinsic variability,  in particular,  radial pulsations (all D-type and some S-type 
systems) and semi-regular variations (S-type) with timescales of order of months and 
years as well as  long-term light variations due to variable obscuration by circumstellar 
dust (most D-type systems), solar-type cycles, spots, etc. The effects of mass accretion 
onto the hot component also involve different variable phenomena. The hot component in 
the vast majority of symbiotic systems seems in fact to be a luminous ($\sim 1000\, \rm 
L_{\sun}$) and hot ($\sim 10^5\, \rm K$) white dwarf powered by thermonuclear burning of 
the material accreted from its companion's wind. Depending on the accretion rate, these 
systems can be either in a steady burning configuration or undergo a hydrogen shell 
flashes. In many cases such flashes can last for decades due to low mass of the white 
dwarf (Miko{\l}ajewska 1997). In addition, the hot components in many systems show 
activity with timescales of a few years which cannot be simply accounted for by the 
thermonuclear models. A possible and promising explanation involves fluctuations in mass 
transfer and/or accretion instabilities.

Below, I present examples of light curves for well-studied, though not yet completely 
understood, symbiotic binaries: RX Pup, CI Cyg and CH Cyg,  which are representative for  
the  wealth  of variable phenomena observed in these systems.

\subsection{RX Puppis: a possible recurrent nova with a  symbiotic Mira companion}

\nobreak

\begin{figure} 
\plotone{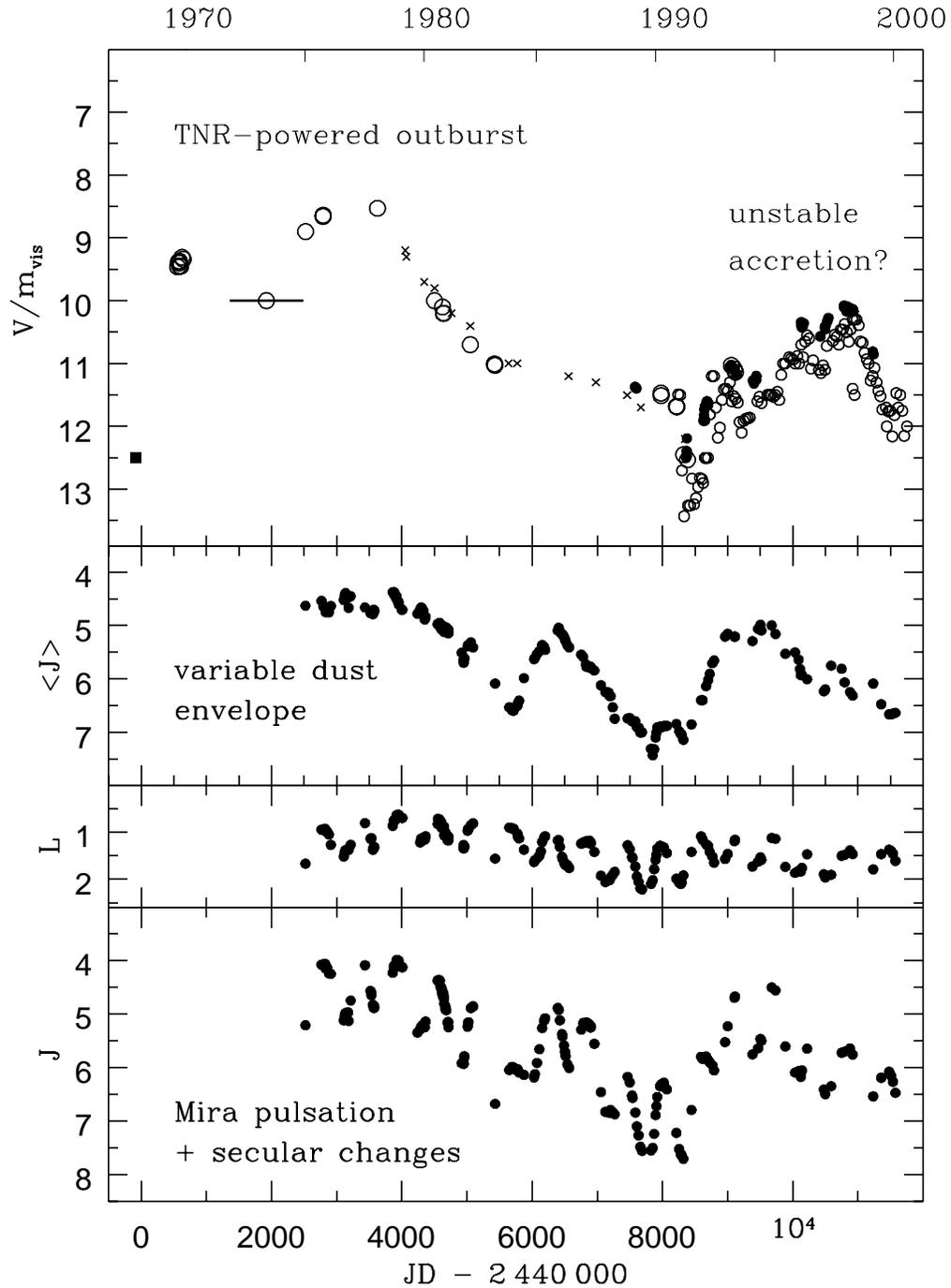} 
\caption{Optical and IR light curves of RX Pup from Miko{\l}ajewska et al. 
(1999).  In the $V/m_{\rm vis}$  light curve, small open circles represent observations 
from RASNZ; large open circles and dots published V magnitudes; crosses FES magnitudes. 
The optical light is dominated by the hot component activity whereas the IR light curves 
are dominated by the Mira pulsation and variable dust obscuration of the Mira by 
circumstellar dust.} 
\end{figure}

RX Pup is a symbiotic binary composed of a long-period Mira variable pulsating with $P 
\approx 578$ days, surrounded by a thick dust shell, and a hot $\sim 0.8\, \rm 
M_{\odot}$, white dwarf companion. The binary separation could be as large as $a \geq 50$ 
a.u. (corresponding to $P_{\rm orb} \geq 200$ yr) as suggested by the permanent presence 
of a dust shell around the Mira component (Miko{\l}ajewska et al. 1999). In particular, 
the Mira is never stripped of its dust envelope,  and even during relatively unobscured 
phases the star resembles the high-mass loss galactic Miras with thick dust shells. In 
general, the binary component separations in D-type systems must be larger than the dust 
formation radius. Assuming a  typical dust formation radius of $\ga 5 \times R_{\rm 
Mira}$, and a Mira radius $R_{\rm Mira} \sim 1 \div 3\, {\rm au}$ (e.g. Haniff, Scholz \& 
Tuthill 1995), the minimum binary separation is $a \ga 20\, {\rm au}$, and the 
corresponding binary period is $P_{\rm orb} \ga 50$\,yr, for {\it any} D-type system.

Recent analysis of multifrequency observations shows that most, if not all, photometric 
and spectroscopic activity of RX~Pup in the UV, optical and radio range is due to 
activity of the hot component, while the Mira variable and its circumstellar environment 
is responsible for practically all changes in the infrared range (Miko{\l}ajewska et al. 
1999, and Fig. 1). In particular, RX~Pup underwent a nova-like eruption during the last 
three decades. The hot component contracted in radius at nearly constant luminosity from 
1975 to 1986, and was the source of strong stellar wind, which prevented it from 
accreting material lost in the Mira wind. Around 1988/9, the hot component turned over 
the 
HR diagram and by 1991 its luminosity had faded by a factor of $\sim 30$ with respect the 
 maximum plateau value (see the very deep minimum in the visual light curve in Fig.1) 
and the hot wind had practically ceased. By 1995 the nova remnant started to accrete 
material from the wind, as indicated by a general increase of the optical flux. The 
earliest observational records from the 1890s suggest that another nova-like eruption of 
RX~Pup occurred around 1894.

The  near-IR light curves show significant long-term variations in addition to the Mira 
pulsation (Fig. 1). The long-term changes are best visible in the $\langle J \rangle$ 
light curve after removal of the Mira pulsation (middle panel in Fig. 1). Miko{\l}ajewska 
et al. (1999) have found large changes in the reddening towards the Mira accompanied by 
fading of the near IR flux. However, the reddening towards the hot component and emission 
line regions remained practically constant and was generally less than that towards the 
Mira. These changes do not seem related to the orbital configuration nor to the hot 
component activity. Similar dust obscuration events seem to occur in many well covered 
symbiotic Miras (e.g. Whitelock 1998), as well as in single  Miras (e.g. Mattei 1997, 
Whitelock 1998), and they are best explained as intrinsic changes in the circumstellar 
environment of the Mira variable, possibly due to intensive and variable mass loss. The 
last increase in extinction towards the Mira in RX Pup has been accompanied by large 
changes in the degree of polarization in the optical and red spectral ranges. This 
confirms  that these long-term variations are driven by changes in the properties of the 
dust grains, such as variable quantity of dust and variable particle size distribution, 
due to dust grain formation and growth (Miko{\l}ajewska 2001).

\subsection{CI Cygni: a tidally distorted giant with a disc-accreting secondary}

Although most symbiotic binaries seem to interact by wind-driven mass loss, a few of them 
may contain a Roche-lobe filling giant. They also show activity with time scales of order 
of years that can be related to the presence of accretion discs. Among them,  CI~Cyg is 
one of the best studied. Kenyon et al. (1991) demonstrated that it consists of an M5 II 
asymptotic branch giant, $M_{\rm g} \sim 1.5\, \rm M_{\sun}$, and a $\sim 0.5\, \rm 
M_{\sun}$ hot companion separated by 2.2 au. They also argued that the hot companion is a 
disc-accreting main sequence star. However, quiescent IUE data from early 1990s can be 
also accounted by a hot and luminous stellar source powered by thermonuclear burning 
which makes the case for CI~Cyg as an accreting MS star less clear.

\begin{figure} 
\plotone{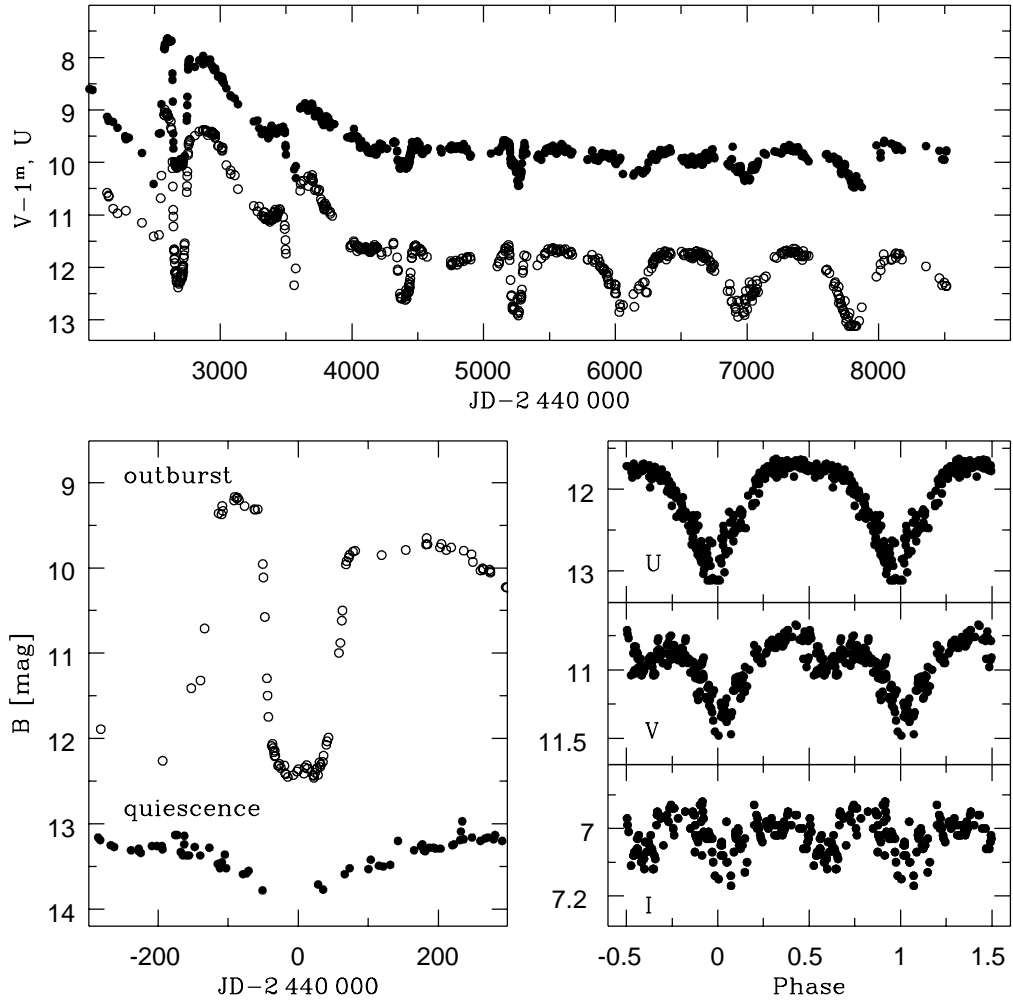} 
\caption{Optical and near infrared light curves of 
CI~Cyg. The $UBVI$ data are from Belyakina (1979, 1984, 1991, 1992), Khudyakova (1989) 
and Meuninger (1981). Deep eclipses and the large outburst which started in 1975 are the 
most prominent features of the optical light curves (upper panel).  During the outburst 
and its decline eclipses were narrow with well-defined eclipse contacts whereas the 
quiescent light curves show very broad minima and almost continuous sinusoidal variation 
(left panel). The ellipsoidal variability of the red giant is visible only in quiescent 
visual and red light curves (right panel).}
\end{figure}

The outburst light curve of CI~Cyg ( Fig. 2) in addition to deep eclipses of the hot 
component by the red giant shows a $0.5-1.0$ oscillations with a period $\sim 0.9\, 
P_{\rm orb}$. Some other S-type systems show similar secondary periodicities best visible 
in their outburst light curves, in all cases $10-15\,\%$ shorter than the orbital 
periods. The nature of this secondary periodicity is unknown. Recently, Galis et al. 
(1999) suggested that in the case of AG Dra, it is due to radial pulsations of the giant, 
and the outbursts are driven by resonances between the pulsations and binary motion. On 
the other hand, the secondary periodicities are best visible in the optical light where 
the contribution from the giant, especially during the outburst, is very low or 
negligible. There is also a striking similarity between these variations and the 
superhumps of the SU UMa class of CVs which may indicate that they are rather related to 
the presence of accretion discs (Miko{\l}ajewska \& Kenyon 1992).

During the outburst and its decline, eclipses in the $UBV$ continuum and optical H\,{\sc 
ii}, He\,{\sc i} and He\,{\sc ii} emission lines were narrow with well-defined eclipse 
contacts whereas at quiescence very broad minima and continuous nearly sinusoidal changes 
are observed. In addition,  the quiescent $VRI$ light curves show a modulation with 
$P_{\rm orb}/2$ as expected for an ellipsoidal light curve. The amplitude of this 
modulation, $\Delta I \sim 0.15$ is consistent with the system  inclination, $i \sim 
73^{\circ}$, and the mass ratio, $M_{\rm g}/M_{\rm h} \sim 3$ derived by Kenyon et al. 
(1991). The transition from narrow eclipses to sinusoidal variations was accompanied by 
large spectral changes and appearance of a radio emission with a spectral distribution 
that cannot be simply accounted for by any of the popular models for symbiotic stars 
(Miko{\l}ajewska \& Ivison 2001).

\subsection{CH Cygni: triple or binary system with a magnetic white dwarf}

\begin{figure}
\plotone{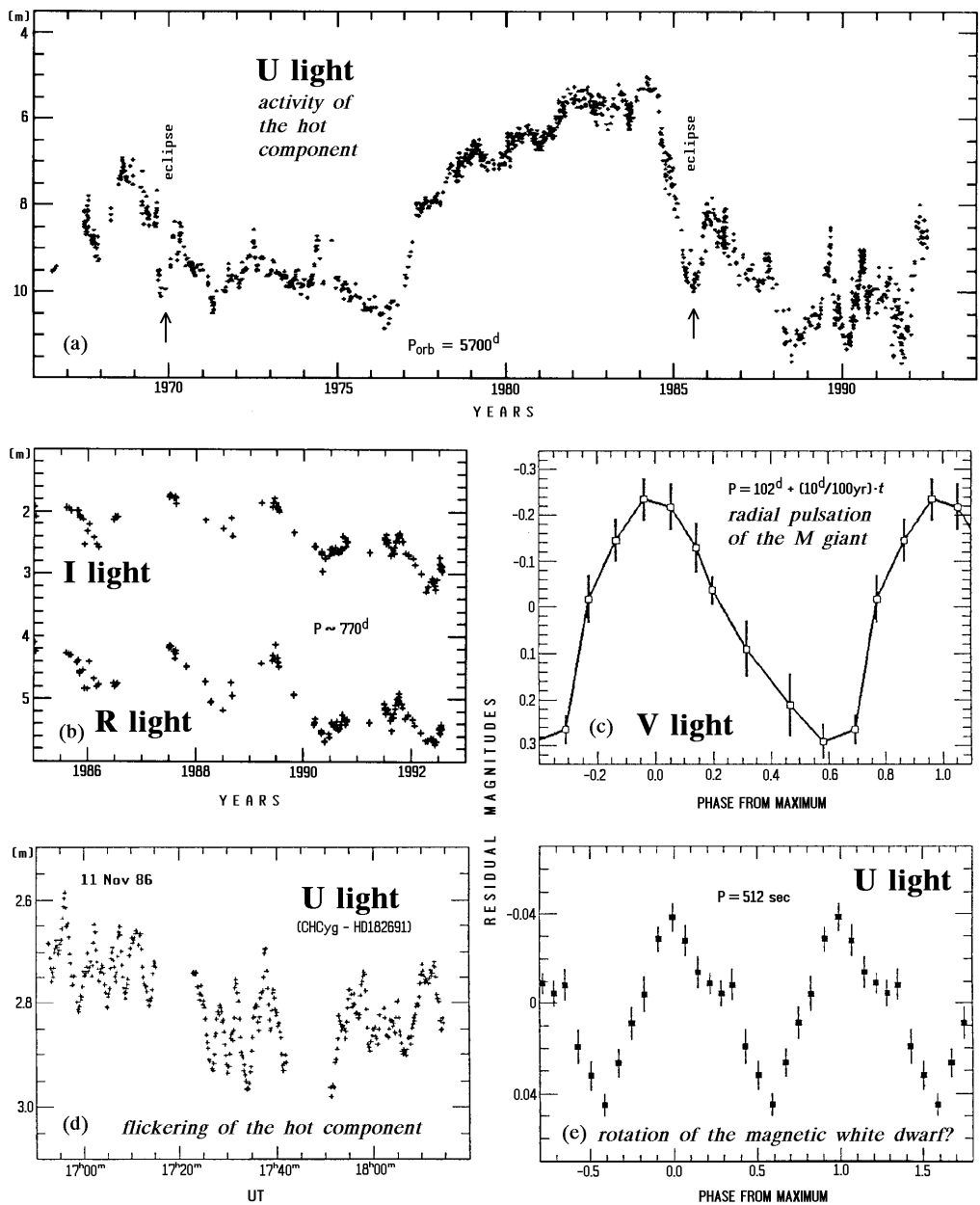}
\caption{Variable phenomena in CH Cyg: (a) U light curve from the recent series of 
outbursts. Bars indicate times  of eclipses of the hot component. (b) R and I light 
curves of the M giant (c) Mean light curve for residuals of the V magnitude binned and 
folded with the 102$^{\rm d}$-pulsation period of the M giant (Miko{\l}ajewski et al. 
1992). (d) Rapid variability in U light. (e) Residual normal points of the data plotted 
in panel (d) folded with the 512-sec period (Miko{\l}ajewski et al. 1990).} 
\end{figure}

The record for the complexity of variable phenomena found in a single symbiotic object 
may be held by CH~Cyg, the symbiotic system with the longest ($P_{\rm orb} \sim 15.5$ yr) 
measured orbital period (Miko{\l}ajewski, Tomov, \& Miko{\l}ajewska 1987; Hinkle et 
al. 1993),  which light curves are presented in Fig. 3. Both the light curves and the 
radial  velocity curves show multiple periodicities: a $\sim 100^{\rm d}$ photometric 
period, best visible in the $VRI$ light curves, has been attributed to radial pulsation 
of the giant (Miko{\l}ajewski, Miko{\l}ajewska, \& Khudyakova 1992), while the nature of 
the secondary period of $\sim 756^{\rm d}$ also present in the radial velocity curve is 
not clear (Hinkle et al. 1993; Munari et al. 1996). There is a controversy about whether 
the system is triple or binary, and whether the symbiotic pair is the inner binary or the 
white dwarf is on the longer orbit. The near-IR light curves also show long-term 
variations similar to the dust obscuration phenomena found in symbiotic Miras (cf. Munari 
et al. 1996).

The hot component also shows very spectacular activity. In particular, we deal with 
irregular outbursts accompanied by fast, massive outflows and jets, rapid brightness 
variations with a time scale of the order of minutes, and other peculiarities which 
cannot be explained in the frame of the classical models proposed for symbiotic stars. 
Miko{\l}ajewski et al. (1990) proposed that this peculiar activity is powered by unstable 
accretion onto a magnetic white dwarf secondary.

\section{Present state-of-the-art  and  future prospects}

Recently published catalogue of symbiotic stars includes 188 symbiotic stars as well as 
30 objects suspected of being symbiotic (Belczy{\'n}ski et al. 2000), Among them, 173 are 
in our Galaxy, 14 in Magellanic Clouds and 1 in Draco-1. They are excellent targets for 
small telescopes, especially for long-term monitoring of their complex photometric and 
spectroscopic variability. Although we have $\sim 120$ S-type symbiotic systems with $V 
\la 15^{\rm m}$, photometric orbital periods have been measured for only 30 objects (18 
of then are eclipsing). 21 systems have also known spectroscopic orbits and for 8 of them 
mass ratios have been  also estimated. The ellipsoidal light variations, characteristic 
of tidally distorted stars, have been rarely observed. Thus far, only four systems, T 
CrB, CI Cyg, BD$-21^{\circ}3873$, and possibly EG And, seem to show such changes. The 
general absence of the tidally distorted giant in symbiotic binaries may be however due 
to the lack of systematic searches for the ellipsoidal variations in the red and near-IR 
range, where the cool giants dominate the continuum light. On the other hand, tidal 
interactions are certainly important in symbiotic systems as suggested by practically 
circular orbits of most ($\sim 80$\, \%) systems with known orbital solutions, and 
specifically of all those showing the multiple CI~Cyg-type  outburst activity. We do not 
know the orbital period for any of the extragalactic symbiotic stars, although 8 of them 
belong to the S-type, and with $V\sim 15-17$ mag, are bright enough  for optical 
monitoring even with a relatively small telescope.

Similarly, among 33 galactic  D-type systems ($K \la 8^{\rm m}$),  pulsation periods have 
been observed -- and thus the Mira presence confirmed -- for only 12 systems. Pulsation 
periods are also unknown for  the few extragalactic D-type systems ($K \sim 10 - 13^{\rm 
m}$). 

Optical and near-IR monitoring of symbiotic stars is essential not only to understand 
variable phenomena  in symbiotic stars and more generally long-period interacting 
binaries, but also to study such phenomena in several other astrophysical environments 
(giant stars, planetary nebulae, novae, supernovae, supersoft X-ray sources, hot stars 
and even AGNs).

Studies of the symbiotic Miras are important for understanding evolution and interaction 
of detached low-mass binaries. For example, there is ample observational evidence for 
systematic differences between the symbiotic Miras and average single galactic Miras. In 
particular, their average pulsation  periods are longer, the colours redder and the 
mass-loss rates higher than typical periods, colours, and mass-loss rates for single 
Miras 
(cf. Miko{\l}ajewska 1999). It is interesting  which and how these differences
are related to the binary nature of symbiotic Miras. The symbiotic Mira  are often 
associated with extended radio and/or optically resolved nebulae. These nebulae have 
usually very complex structure, often with bipolar lobes and jet-like features.

There are also many important questions posed by the active S-type systems. What powers 
the multiple outburst activity in CI~Cyg and other similar systems? How many of these 
contain tidally distorted giants? What is the nature of the secondary periodicity, $\la 
0.8-0.9\,P_{\rm orb}$, visible at outburst in some of them? Can the secondary periodicities 
be considered as an evidence for the presence of an accretion disc? Such 
periodicities have not been found in any symbiotic nova during both optical maximum and 
constant luminosity phase (the plateau portion of white dwarf cooling tracks), 
including the best studied case -- AG ~Peg (Kenyon et al. 1993), and 
their 
presence indicates that the outbursts in CI~Cyg and other similar systems are not 
powered by thermonuclear reactions. The timescales and relative amplitudes for these 
eruptions are very similar to the timescales and amplitudes of the hot component 
luminosity variations (high and low states) in symbiotic recurrent novae (e.g. T CrB, RS 
Oph, RX Pup) between their nova eruptions (Anupama \& Miko{\l}ajewska 1999; 
Miko{\l}ajewska et al. 1999) and in other accretion-powered systems (CH Cyg, Mira A+B). 
It is possible that the main difference between CI Cyg, Z And, AX Per, and other 
related symbiotic systems with multiple eruption activity and the activity of 
accretion-powered systems (symbiotic recurrent novae and CH Cyg) is that the hot 
component in the former burns more or less stably the accreted hydrogen whereas not in 
the latter.
Systematic optical and near infrared monitoring with small telescopes can 
provide an answer to these and many other questions.

\acknowledgments

I gratefully acknowledge Maciej Miko{\l}ajewski and Toma Tomov for providing Figure 3. I 
would also like to thank the LOC for their support. This research was partly founded by 
KBN Research Grant No. 5\,P03D\,019\,20.

\end{document}